\documentstyle{article}
\begin{document}
{%\large
\baselineskip=8mm
\begin{center}
{\Large \bf ACOUSTIC PERTURBATIONS IN SPECIAL-RELATIVISTIC PARALLEL FLOWS}\\
\vskip1.5cm
{ANDRIA D. ROGAVA}\\
\vskip0.5cm
{\it Department of Theoretical Astrophysics,  Abastumani  Astrophysical
Observatory, Republic of Georgia, Department of Physics, Tbilisi State
University, Republic of Georgia, and International Center of Theoretical
Physics, Trieste, Italy}\\
\vskip 0.5cm
{VAZHA I. BEREZHIANI}\\
\vskip0.5cm
{\it International Centre for Theoretical Physics, Trieste, Italy and
Department of Plasma Physics,  Institute of Physics, Tbilisi, Republic
of Georgia}\\
\vskip0.5cm
{AND}\\
\vskip 0.5cm
{SWADESH M. MAHAJAN}\\
\vskip 0.5cm
{\it Institute for Fusion Studies, The University of Texas at Austin,
Austin, Texas, USA and International Centre for Theoretical Physics,
Trieste, Italy}
\end{center}
\date{\today}
\vskip 2cm
\newpage
\begin{abstract}
Acoustic perturbations in a parallel relativistic flow of an inviscid fluid
are considered. The general expression for the frequency of the sound waves
in a {\it uniformly} (with zero shear) moving medium is derived.
It is shown that relativity evokes a difference in the frequencies of the
sound-type perturbations propagating  along and against the
current. Besides, it is shown that the perturbations {\it are not} purely
irrotational as they are in nonrelativistic case.
For a
{\it non-uniformly} (with nonzero shear) moving fluid  a general set of
equations, describing the evolution of the acoustic perturbations in
relativistic sheared flows, is obtained and analysed when the
temperature is nonrelativistic. It is shown that, like the nonrelativistic
 case, in the new system:
(a) the excitation of vortical, transiently growing perturbations, and (b)
the excitation of sound-type perturbations, extracting the kinetic energy
of the background flow, are possible. It is demonstrated that the
relativistic character of the motion significantly intensifies the
efficiency of both   these processes. Finally, the possible relevance of
this study to processes in astrophysical shear flows is pointed out.
\end{abstract}
\vskip 2cm

\section{Introduction}

Acoustic phenomena are abundant in Nature. Compressibility of all known
kinds of continuous media implies the possibility of excitation and
propagation of different kinds of acoustic perturbations. In the neutral
fluids and gases, we have the plain sound waves (Lighthill 1978), in
magnetohydrodynamical (MHD) fluids and plasmas, the magnetosonic, and
ion-sound waves (Sturrock 1994), and in complex media like the dusty
plasmas we can excite dust-acoustic and dust ion-acoustic modes (Rao, Shukla
\& Yu 1990, Shukla \& Silin 1992).

The nature of all these compressional modes can be seriously affected in
differentially moving media (shear flows). For example, the perturbations
can show large transient growths and extract energy from the mean flow
(Chagelishvili, Rogava \& Segal 1994). In a variety of physical situations,
where more
than one kind of oscillation mode is sustainable, the shear flow also allows
a mechanism of mutual transformation of the modes and hence of mutual
exchange of energy between them (Chagelishvili, Rogava \& Tsiklauri 1996,
Rogava, Mahajan \& Berezhiani 1996).

The purpose of this paper is to examine the velocity shear induced acoustic
processes for {\it relativistically} moving media. This problem, over and
above
its purely theoretical significance, may have applications to a number of
astrophysical shear flows [e.g., pulsar magnetospheres (Arons \& Smith
1979), bipolar and unipolar outflows in Active Galactic Nuclei
(AGN) (Blandford \& Rees 1974)],
where {\it relativistic shear flows} actually
occur. We shall ``introduce" special-relativistic dynamical effects for the
 simplest, model parallel flow with linear shear; the idea is to single
out the principal novelty evoked by the relativistic character of the
motion. It is hoped that this simple analysis will provide a foundation for
a more thorough treatment of realistic astrophysical shear flows, which are
often rather complicated.

The paper is organized in the following way. In sec.2 we write down
the general system of equations of the special-relativistic fluid mechanics.
We also consider the regular motion of the idealized, plane, parallel
"shearing sheet" of the viscous fluid, and derive an analytic expression for
the mean (equilibrium) velocity profile.

Section 3 deals with the linear theory of the acoustic perturbations in this
flow. We first study a uniform (shearless) relativistic flow for which a
general expression for the frequency of the sound waves is derived. It is
shown that when both the mean velocity of the flow and the temperature of
the fluid are relativistic, there appears the remarkable (though expected)
difference in the frequencies of the sound waves propagating down the
current and against the current.
Another interesting feature of the sound-type perturbations is that they are
not curl-free as they are in nonrelativistic hydrodynamics.
We then introduce shear, and derive
the general system of equations describing the temporal evolution of the
acoustic perturbations. The particular case of the relativistic shear flow
with nonrelativistic temperature is analyzed in detail. The shear induced
processes in the flow are rather similar to their nonrelativistic
(Chagelishvili, Rogava \& Segal 1994)
analogues: we find almost incompressible, vortex (vortical) perturbations
that exhibit nonexponential, algebraic transient growth, and the usual,
compressible (sound-type) perturbations that are able to extract the
kinetic energy of the mean flow. It is shown that the rates of the algebraic
growth, and also of energy exchange are enhanced by a factor
${\gamma}_0^2$ (${\gamma}_0$ being the average Lorentz factor of the flow)
resulting in tremendous efficiency for extracting energy from the
superrelativistic shear flows.

In the conclusion we briefly discuss the possible areas of astrophysical
interest where the results of this study can be applied.

\section{Governing equations: mean flow dynamics}

Throughout this paper, we shall use the following notation:~ (a) The
Greek
indices will range over $t,~x,~y,~z$ and represent space-time coordinates,
components, etc.; (b)~The Latin indices will range over $x,~y,~z$ and
represent coordinates in the three-dimensional space. We use
geometrical units, so that the speed of light $c=1$.

The spacetime metric is Minkowskian:
$$
ds^2=-d{\tau}^2={\eta}_{{\alpha}{\beta}}dx^{\alpha}dx^{\beta}=
-dt^2+dx^2+dy^2+dz^2, \eqno(1)
$$
and the stress-energy tensor for a relativistic, viscous, neutral fluid
consisting of particles of $m_0$, may be written as
$$
T^{{\alpha}{\beta}}{\equiv}(e+P)U^{\alpha}U^{\beta}+
P{\eta}^{{\alpha}{\beta}}-2{\eta}{\sigma}^{{\alpha}{\beta}}-
{\xi}{\theta}h^{{\alpha}{\beta}}, \eqno(2)
$$
in the standard notation. Here $U^{\alpha}$ is the four velocity
vector with components:
$$
U^{\alpha}=[{\gamma},~~{\gamma}
{\bf V}^i],~~~~~~~~{\gamma}{\equiv}(1-{\bf V}^2)^{-1/2}, \eqno(3)
$$
$e$ is the specific (proper) energy density, $P$ is the specific
pressure and $\eta$, $\xi$ are the coefficients of specific shear and bulk
viscosity respectively. The {\it projection tensor} $h^{{\alpha}{\beta}}$ is
defined as:
$$
h^{{\alpha}{\beta}}{\equiv}{\eta}^{{\alpha}{\beta}}
+U^{\alpha}U^{\beta}, \eqno(4)
$$
while $\theta$ is the "expansion" of the fluid world lines:
$$
\theta{\equiv}{U^{\alpha}}_{;\alpha}, \eqno(5)
$$
and ${\sigma}^{{\alpha}{\beta}}$ is the shear tensor:
$$
{\sigma}_{{\alpha}{\beta}}{\equiv}{\left[U_{{\alpha};{\mu}}
h^{\mu}_{\beta}+U_{{\beta};{\mu}}h^{\mu}_{\alpha} \right]}/2-
({\theta}/3)h_{{\alpha}{\beta}}. \eqno(6)
$$

Note that a thermodynamic relation between $e$ and $P$ implies:
$$
e={\rho}+({\Gamma}-1)^{-1}P, \eqno(7)
$$
where ${\rho}{\equiv}m_0n$ is the proper material density (rest mass
density) of the fluid, and
$n$ is the proper number density of particles. Note also that ${\Gamma}=5/3$
for the fluid with nonrelativistic temperature and ${\Gamma}=4/3$ for
the one with ultrarelativistic temperature.

The equations of motion of the dissipative fluid are contained in the
conservation law of the stress-energy tensor,
$$
{T_{\alpha}}^{\beta}_{;\beta}=0. \eqno(8)
$$

Let us first consider the equilibrium; we have a relativistic, idealized,
plane,
parallel "shearing sheet" of the viscous fluid. The fluid moves in the x-th
direction, its lower layer ($y=0$) moves with a relativistic velocity
$V_0{\le}1$, while the upper one ($y=L$) moves with $V_0+{\Delta}V$
(${\Delta}V{\ll}V_0$), $L$ being the width of the "shearing sheet."
The symmetry of the problem implies that the nonzero component
of the fluid {\it mean} 3-velocity ${\bar V}_x$ can depend only on the $y$
coordinate. The nonzero components of the mean 4-velocity are:
$$
U^t=(1-{\bar V}_x^2)^{-1/2}{\equiv}{\bar{\gamma}}, \eqno(9a)
$$
$$
U^x={\bar{\gamma}}{\bar V}_x, \eqno(9b)
$$
Equations (9a) and (9b), coupled with the x-th component of the
stress-energy conservation equation (8) yield:
$$
{{{\partial}^2U^x}\over{{\partial}y^2}}=0, \eqno(10)
$$
with the general solution, $U^x=Ay+B$, which immediately
leads to the following {\it nonlinear} mean velocity profile:
$$
{\bar V}_x(y)={{Ay+B}\over{\sqrt{1+(Ay+B)^2}}}, \eqno(11)
$$
where $A$ and $B$ are the constants of integration, to be determined through
the boundary conditions [${\gamma}_0=(1-V_0^2)^{-1/2}$]:
$$
B={\gamma}_0V_0, \eqno(12a)
$$
$$
A={1 \over  L}{\left[{{V_0+{\Delta}V}\over{\sqrt{1-(V_0+{\Delta}V)^2}}}-
{\gamma}_0V_0\right]}. \eqno(12b)
$$

When ${\Delta}V{\ll}V_0$, the last expression leads to the simple result:
$$
A{\simeq}{\gamma}_0^3({\Delta}V/L). \eqno(13)
$$

>From now on we shall assume that $AL{\ll}B$, which holds if
${\Delta}V{\gamma}_0^2{\ll}V_0$. In this
case (11) leads to the following approximate expressions [$a{\equiv}
{\Delta}V/L$]:
$$
{\bar V}_x(y){\simeq}V_0+ay, \eqno(14)
$$
for the mean flow velocity, and
$$
{\bar{\gamma}}(y){\simeq}{\gamma}_0(1+aV_0{\gamma}_0^2y). \eqno(15)
$$
for the corresponding Lorentz factor.

\section{Perturbations: linear treatment}

After having established an equilibrium, we now proceed to examine the
response of the mean flow to linear perturbations. For simplicity, we
neglect the viscous terms in the stress-energy tensor. The relevant equations
are the {\it particle number conservation}
$$
(nU^{\alpha})_{,{\alpha}}=0, \eqno(16)
$$
which in the usual vector notation reads:
$$
{\partial}_t(n{\gamma})+{\nabla}{\cdot}(n{\gamma}{\bf V})=0,
\eqno(17)
$$
and the equation for {\it momentum conservation}. The latter is conveniently
derived by
means of the projection tensor $h^{{\alpha}{\beta}}$, operating on the
stress-energy conservation. The required equation
$$
h_i^{\beta}{T_{\beta}^{\alpha}}{,{\alpha}}=
(e+P)U^{\alpha}U_{i,{\alpha}}+U_iU^{\alpha}P,{\alpha}+P_{,i}=0, \eqno(18)
$$
may be written as:
$$
(e+P){\gamma}D_t({\gamma}{\bf V})+{\gamma}^2{\bf V}D_tP+{\nabla}P=0,
\eqno(19)
$$
in the more familiar vector notation. In Eq.(19) $D_t{\equiv}[{\partial}_t+
({\bf V},{\nabla})]$ is the {\it convective derivative}.

For further analysis it is more convenient to write (17) in the following way
$$
D_t{\varrho}+{\varrho}({\nabla}{\cdot}{\bf V})=0, \eqno(20)
$$
where ${\varrho}{\equiv}m_0n{\gamma}={\rho}{\gamma}$ is the fluid mean
density in the laboratory frame.

Note that due to (15), the mean value of ${\varrho}$ is
spatially inhomogeneous $[{\bar{\rho}}=m_0{\bar n}]$:
$$
{\bar{\varrho}}={\bar{\rho}}{\gamma}_0(1+aV_0{\gamma}_0^2y). \eqno(21)
$$

The inhomogeneity of the mean flow is the fundamental object of our current
investigation. Our later calculations show how relativity ``creates"
enormous effective inhomogeneities as seen in the laboratory frame. We
present here a basic physical picture of the expected phenomenon: Let us
assume that the fluid is {\it inherently homogeneous}, i.e., its proper
mean particle number density ${\bar n}$, and its specific mean
rest mass density
${\bar {\rho}}$ are constants. If we imagine this fluid to be consisting
of different ``layers," then while at rest, the number of ``fluid
elements" per unit ``length" of the layer, as seen in the laboratory frame,
is the same in {\it all} layers. In other words the density ${\bar
n}=const$. For a differentially moving fluid, i.e., when
the layers have {\it different} velocities in the laboratory frame, the
situation is much more interesting. Due to the ``Lorentz
contraction," the distances between adjacent fluid particles in each
layer (as seen in laboratory frame) will decrease; the higher the
velocity, the greater the contraction. Thus we arrive at the important
conclusion:
{\it The relativistic differential motion of the fluid induces a density
stratification (inhomogeneity) in the originally homogeneous fluid}. It
seems reasonable to expect that this {\it special-relativistic effect}
should play a crucial role in physical processes dominated by
inhomogeneities.

Let us now decompose the instantaneous values of all physical variables
appearing in (19) and (20), into their perturbational and mean components:
$$
{\varrho}={\bar{\varrho}}+{\varrho}{'}, \eqno(22a)
$$
$$
{\bf V}={\bar {\bf V}}+{\bf v}, \eqno(22b)
$$
$$
{\gamma}={\bar{\gamma}}+{\gamma}^{'}={\gamma}_0(1+aV_0{\gamma}_0^2y+
V_0{\gamma}_0^2v_x). \eqno(22c)
$$

Using (22a)--(22c), the linearized evolution equations take the form
$$
[{\partial}_t+(V_0+ay){\partial}_x]{\varrho}^{'}+{\bar{\rho}}{\gamma}_0
({\partial}_xv_x+{\partial}_yv_y)+aV_0{\gamma}_0^3{\rho}v_y=0, \eqno(23)
$$
$$
[{\partial}_t+(V_0+ay){\partial}_x]v_x +av_y+{{V_0}\over{({\bar e}+{\bar P})
{\gamma}_0^2}}[{\partial}_t+(V_0+ay){\partial}_x]P^{'}+
{1\over{({\bar e}+{\bar P}){\gamma}_0^4}}{\partial}_xP^{'}=0, \eqno(24)
$$
$$
[{\partial}_t+(V_0+ay){\partial}_x]v_y+{1\over{({\bar e}+{\bar P})
{\gamma}_0^2}}{\partial}_yP^{'}=0. \eqno(25)
$$
Note that the term $aV_0{\gamma}_0^3{\rho}v_y$  in (23) arises due to the
above described relativistic density stratification effect.

In writing (23--25), we have neglected the y-dependent parts in the
coefficients of certain terms. However, we do retain the $ay$ term in the
convective derivative. This term representing shear, even when
$ay{\ll}V_0$, is the source of much of the interesting physics discussed in
this paper. Note that in (23) ${\varrho}^{'}$  is the density perturbation 
as measured in the laboratory frame, while $P^{'}$ in (24) and (25) 
is the specific pressure perturbation as measured in the fluid rest frame.
It is more convenient to express ${\varrho}^{'}$ through the corresponding 
specific density perturbation in order to have in equations all thermodynamic
quantities relative to the one and the same frame of reference. The
result is:
$$
{\varrho}^{'}={\bar{\gamma}}{\rho}^{'}+{\gamma}^{'}{\bar {\rho}}
{\simeq}{\gamma}_0({\rho}^{'}+{\bar {\rho}}V_0{\gamma}_0^2v_x).
\eqno(26)
$$

Defining $d{\equiv}{\rho}^{'}/{\bar {\rho}}$ we can rewrite (23) as
$$
[{\partial}_t+(V_0+ay){\partial}_x](d+V_0{\gamma}_0^2v_x )+{\partial}_xv_x
+{\partial}_yv_y+aV_0{\gamma}_0^2v_y=0. \eqno(27)
$$

The combination $({\bar e}+{\bar P})$ has a simple physical meaning in either
of the extreme limits of the nonrelativistic and the ultrarelativistic
temperatures.
In the former case $({\bar e}+{\bar P}){\simeq}{\bar {\rho}}$, while in the
latter one $({\bar e}+{\bar P}){\simeq}4{\bar P}$. In both cases the term
is expressible through the sound speed $C_s$. For nonrelativistic
temperatures $C_s{\equiv}({\partial}P/{\partial}{\rho})^{1/2}{\ll}1$, while
in the case of the ultrarelativistic temperatures $C_s=1/{\sqrt{3}}$.

After making the change of variables
$$
t_1=t;~~x_1=x-(V_0+ay)t;~~y_1=y, \eqno(28)
$$
which leads to the following transformation of operators
$$
[{\partial}_t+(V_0+ay){\partial}_x]={\partial}_{t_1};~~{\partial}_x=
{\partial}_{x_1};~~{\partial}_y={\partial}_{y_1}-At_1{\partial}_{x_1},
\eqno(29)
$$
we can write the closed set of  differential equations for three
variables: $v_x$, $v_y$ and $d$:
$$
{\partial}_{t_1}(d+V_0{\gamma}_0^2v_x )+{\partial}_{x_1}v_x
+({\partial}_{y_1}-At_1{\partial}_{x_1})v_y+aV_0{\gamma}_0^2v_y=0, \eqno(30)
$$
$$
{\partial}_{t_1}v_x +av_y+(V_0C_s^2/{\gamma}_0^2){\partial}_{t_1}d+
(C_s^2/{\gamma}_0^4){\partial}_{x_1}d=0, \eqno(31)
$$
$$
{\partial}_{t_1}v_y+(C_s^2/{\gamma}_0^2)
({\partial}_{y_1}-At_1{\partial}_{x_1})d=0. \eqno(32)
$$

The set of Eqs. (30--32) are now Fourier analyzed in $x_1$ and $y_1$ through
the definition
$$
F=\int{{\hat F}({{k_x}_1},{{k_y}_1},t_1)
exp[i({{k_x}_1}x_{1}+{{k_y}_1}y_{1})]}d{{k_x}_1}d{{k_y}_1}. \eqno(33)
$$
Introducing dimensionless variables: (${\tau}{\equiv}k_{x_1}t_1$,
$R{\equiv}a/k_{x_1}$, ${\beta}_0{\equiv}k_{y_1}/k_{x_1}$,
${\beta}(\tau)={\beta}_0-R{\tau}$), we can reduce the Fourier transformed
system to the following closed set of first order, ordinary
differential equations (ODE's) [$C^2{\equiv}C_s^2/{\gamma}_0^2$]:
$$
{\partial}_{\tau }(d+V_0{\gamma}_0^2v_x )+i[v_x
+{\beta}(\tau)v_y]+RV_0{\gamma}_0^2v_y=0, \eqno(34)
$$
$$
{\partial}_{\tau}v_x +Rv_y+V_0C^2{\partial}_{\tau}d+
i(C^2/{\gamma}_0^2)d=0, \eqno(35)
$$

$$
{\partial}_{\tau}v_y+iC^2{\beta}(\tau)d=0. \eqno(36)
$$

In the nonrelativistic limit ($V_0{\ll}1$, ${\gamma}_0{\simeq}1$)
these equations reduce to the system that was analyzed
by Chagelishvili, Rogava \& Segal (1994).

Multiplying (35) by $V_0{\gamma}_0^2$ and combining it with (34), we can
write, instead of these two equations, a simple pair
$$
{\delta}{\partial}_{\tau}d=i{\left[V_0C^2d-(v_x+{\beta}v_y)\right]}, \eqno(37)
$$
$$
{\delta}{\partial}_{\tau}v_x=-{\delta}Rv_y-i(C^2/{\gamma}_0^2)d+
iV_0C^2(v_x+{\beta}v_y). \eqno(38)
$$
where ${\delta}{\equiv}1-C_s^2V_0^2$, is limited to the range
$2/3<{\delta}<1$, and differs noticeably from 1 only for the case of
ultrarelativistic temperatures ($C_s^2=1/3$) coupled with  highly
relativistic background flow velocity
($V_0{\simeq}1$). In all other cases the parameter is close to 1.

From (36) and (38) one can easily deduce
$$
{\partial}_{\tau}[v_y-{\delta}{\gamma}_0^2{\beta}v_x]={\gamma}_0^2
({\delta}R-iV_0C^2{\beta})[v_x+{\beta}v_y]. \eqno(39)
$$
This relation will be useful in the following considerations. It is also
useful to know that the parameter $R{\sim}{\Delta}V({\ell}_x/L){\ll}1$
(${\ell}_x$ is a characteristic length scale of the perturbation along X)
because both ${\Delta}V$ and ${\ell}_x/L$ are much less than unity.

\subsection{Zero shear ($R=0$) case}

We first investigate the shearless ($R=0$) system. For this case all
coefficients in (36) and (38--39) become constants [${\beta}(\tau)=
{\beta}_0=const$] and the following dispersion relation
$$
\left(\matrix{
{\delta}{\Omega}-V_0C^2&          1            &   {\beta}_0           \cr
C^2/{\gamma}_0^2       &{\delta}{\Omega}-V_0C^2&   -V_0C^2{\beta}_0    \cr
C^2{\beta}_0           &          0            &   {\Omega}            \cr
}\right)=0,
\eqno(40)
$$

which can be expanded to yield
$$
{\Omega}({\delta}{\Omega}-V_0C^2)^2-{\Omega}C^2(1/{\gamma}_0^2+{\delta}
{\beta}_0^2)=0,
\eqno(41)
$$
gives the normal-mode frequencies,
$$
{\Omega}_{\pm}={1 \over {\delta}}{\left[V_0C^2{\pm}
C{\sqrt{{1 \over{{\gamma}_0^2}}+{\delta}{\beta}_0^2}}~\right]}. \eqno(42)
$$

Before analysing (42), it is interesting to show that the dispersion
relation could be obtained by a Lorentz transformation of the sound
wave dispersion in the rest frame of the fluid,

$$
{\hat {\omega}}^2=C_s^2({\hat k}_x^2+{\hat k}_y^2). \eqno(43)
$$
Now the four-vectors $({\omega},~{\bf k})$ in the laboratory frame, and
$({\hat{\omega}},~{\hat{\bf k}})$ in the rest frame are the Lorentz
transforms of one another,
$$
{\hat {\omega}}={\gamma}_0({\omega}+V_0k_x), \eqno(44a)
$$
$$
{\hat k}_x={\gamma}_0(k_x-V_0{\omega}), \eqno(44b)
$$
$$
{\hat k}_y=k_y. \eqno(44c)
$$

The sound frequency ${\Omega}$, associated with the coordinates
($t_1,~x_1,~y_1$), is related to the oscillation frequency $\omega$ by
$$
{\omega}={\Omega}-V_0k_x. \eqno(45)
$$
and the corresponding phase is
${\varphi}{\equiv}[{\Omega}t_1+k_{x_1}x_1+k_{y_1}y_1]=[({\Omega}-V_0k_x)
t+{k_x}x+{k_y}y]$, since $k_{x_1}=k_x$ and $k_{y_1}=k_y$ in the shearless
case.

Inserting (44) and (45) in (43) we get the following quadratic equation
$$
(1-C_s^2V_0^2){\Omega}^2-{{2V_0C_s^2}\over{{\gamma}_0^2}}k_x{\Omega}-
{{C_s^2}\over{{\gamma}_0^2}}{\left({{{k_x}^2}\over{{{\gamma}_0}^2}}+
k_y^2\right)}=0, \eqno(46)
$$
whose solution is
$$
{\Omega}_{\pm}={1 \over {{\delta}{\gamma}_0^2}}{\left[V_0C_s^2k_x{\pm}
C_s{\sqrt{k_x^2+{\delta}{\gamma}_0^2k_y^2}
}~\right]}, \eqno(47)
$$
which is exactly the same as (42); only the normalization is different.

Let us now examine Eq.(42) in some detail:

\begin{itemize}
\item
In the nonrelativistic limit ($V_0{\ll}1$, ${\gamma}_0{\simeq}1$) it reduces
to the usual dispersion relation describing sound waves:
$$
{\Omega}_{\pm}={\pm}C_s{\sqrt{1+{\beta}_0^2}}. \eqno(48)
$$

\item
for ultrarelativistic temperatures and nonrelativistic mean
motion the same relation (48) is valid, but now the speed of sound attains
its supreme value: $C_s=1/{\sqrt{3}}$.

\item
When the flow mean velocity is relativistic, but the temperature is
nonrelativistic the first term in the brackets in (42) is still much
smaller than 1 and, therefore, the expression for the sound wave
frequencies reads:
$$
{\Omega}_{\pm}={\pm}(1-V_0^2)C_s{\sqrt{1+
{\beta}_0^2/(1-V_0^2)}}. \eqno(49)
$$

\end{itemize}

In all these cases the absolute values of the frequencies, describing the
sound waves propagating along (${\Omega}_{+}$) and against (${\Omega}_{-}$)
the background flow, respectively, are equal. The degeneracy is, however,
destroyed for ultrarelativistic temperatures (when $C_s^2$ is not too small)
and highly relativistic values of the average flow velocity $V_0$ of the
shearing sheet; the two frequencies, as it is clear from (42), could be
vastly different.

Another interesting feature of the sound waves in a special-relativistic,
uniform flow is that sound-type perturbations {\it are not} purely
irrotational (curl-free) as they are in the standard nonrelativistic
hydrodynamics. We can see this by examining the $R=0$ limit of Eqs. (34)
and (39),
$$
{\partial}_{\tau }(d+V_0{\gamma}_0^2v_x )=-i[v_x+{\beta}_0v_y], \eqno(50)
$$
$$
{\partial}_{\tau}[v_y-{\delta}{\gamma}_0^2{\beta}_0v_x]=-iV_0{\gamma}_0^2C_s^2
{\beta}_0[v_x+{\beta}_0v_y], \eqno(51)
$$
which immediately lead to the algebraic relation,
$$
v_y-{\gamma}_0^2{\beta}_0v_x=V_0C_s^2{\beta}_0d+{\it const}. \eqno(52)
$$

In the nonrelativistic limit, this relation reduces to the
conventional condition $v_y-{\beta}_0v_x={\it const}$, which, for
sound-type perturbations (${\it const}=0$) implies irrotationality of these
perturbations [~$(v_y-{\beta}_0v_x){\sim}({\bf k}{\times}{\bf v})_z=0$].
Generally, equation (52), however, does not permit such a conclusion.

\subsection{Non Zero Shear}

To discuss the nonzero shear case, it is convenient to combine
Eqs.(36)--(38) to obtain a single  differential equation. By straightforward
algebra, we can eliminate $v_x$ and $v_y$ to derive an ODE of the third
order in the variable $d$,
$$
{\delta}{\partial}^3_{\tau}d-2iV_0C^2{\partial}^2_{\tau}d+C^2{\left({1\over
{{\gamma}_0^2}}+{\beta}^2\right)}{\partial}_{\tau}d-4RC^2{\beta}d=0. \eqno(53)
$$

The other physical variables ($v_x$ and $v_y$) can be determined in terms of
$d(\tau)$, and its first and second derivatives:
$$
v_y(\tau)=-{1 \over {2R}}{\left[i{\delta}{\partial}^2_{\tau}d+2V_0
C^2{\partial}_{\tau}d+iC^2{\left({1\over{{\gamma}_0^2}}+{\beta}^2\right)}
d\right]}, \eqno(54)
$$
$$
v_x(\tau)=i{\delta}{\partial}_{\tau}d+V_0C^2d-{\beta}v_y. \eqno(55)
$$

A full analysis of the problem will require a numerical study of (53), and
will be attempted at a later stage. To appreciate the new effects introduced
by relativity, let us consider a strongly relativistic ($V_0{\simeq}1$,
${\gamma}_0^2{\gg}1$) shearing sheet with nonrelativistic ($C_s^2{\ll}1$)
temperature. For this case, the defining equations (37)--(39) can be
approximated by
$$
{\partial}_{\tau}d=-i(v_x+{\beta}v_y), \eqno(56)
$$
$$
{\partial}_{\tau}v_x=-Rv_y-i(C_s^2/{\gamma}_0^4)d. \eqno(57)
$$
and
$$
{\partial}_{\tau}[v_y-{\gamma}_0^2{\beta}v_x]=R{\gamma}_0^2
[v_x+{\beta}v_y], \eqno(58)
$$

It should be noted that the terms proportional
to $V_0C_s^2/{\gamma}_0^2$ are
negligible even in the ultimate case of ultrarelativistic temperatures and
superrelativistic mean velocity. This coefficient attains its maximum value
of $2{\sqrt{3}}/27{\simeq}0.1283$ when $V_0=C_s=1/{\sqrt{3}}$.

Combining (56)--(58) we can get the following algebraic relation between
the physical variables:
$$
v_y-{\gamma}_0^2{\beta}v_x=iR{\gamma}_0^2d+const, \eqno(59)
$$

and derive the following second order
inhomogeneous differential equation for $v_x(\tau)$
$$
{\partial}^2_{\tau}v_x+{{C_s^2}\over{{\gamma}_0^4}}(1+{\gamma}_0^2{\beta}^2)
v_x+const{\times}{{C_s^2}\over{{\gamma}_0^4}}{\beta}=0, \eqno(60)
$$

The general solution of (60) is the obvious
sum of its {\it special} solution, and the  appropriate solution of the
corresponding homogeneous equation: $v_x={\bar v}_x+{\hat v}_x$.
When ${\Omega}(\tau){\equiv}(C_s/{\gamma}_0^2)
[1+{\gamma}_0^2{\beta}^2]^{1/2}$ has a slow
dependence on time ${\tau}$, i.e.,
$$
|{{\Omega}(\tau)}^{(1)}|{\ll}{\Omega}^2(\tau), \eqno(61)
$$
it becomes possible to obtain approximate solution to the homogeneous
equation.

For flows with $R{\ll}1$, condition (61) holds for a wide
range of possible values of $|{\beta}(\tau)|$, implying that the WKB solution
$$
{\bar v}_x(\tau){\approx}{C \over{\sqrt{{\Omega}(\tau)}}}exp
[i({\varphi}(\tau)+{\varphi}_0)], \eqno(62)
$$
will be valid for the required range of ${\tau}$. In Eq.(62), the phase
$$
{\varphi}(\tau)={\int}{\Omega}(\tau)d{\tau}=-
{{C_s} \over {2R{\gamma}_0^2}}{\left[{\beta}
{\sqrt{1+{\gamma}_0^2{\beta}^2}}+{1 \over {\gamma}_0}ln{\left|{\gamma}_0
{\beta}+{\sqrt{1+{\gamma}_0^2{\beta}^2}}\right|}\right]}. \eqno(63)
$$

Due to the smallness of $R$, the particular solution of (60) can be sought
through the following series (Magnus, 1976):
$$
{\hat v}_x(\tau)=const{\times}\sum_{n=0}^{\infty}R^{2n}y_n(\tau), \eqno(64)
$$
$$
y_0(\tau)=-{\beta}(\tau)/(1+{\gamma}_0^2{\beta}^2), \eqno (65a)
$$
$$
y_n(\tau)=-{1 \over {\Omega}^2(\tau)}{{{\partial}^2y_{n-1}}
\over {{\partial}{\beta}^2}}. \eqno (65b)
$$

Retaining only the lowest order term in $R$, the approximate solution of the
inhomogeneous equation (60) may be written as:
$$
v_x={\bar v}_x+{\hat v}_x{\approx}{C \over {\sqrt{{\Omega}(\tau)}}}
exp[i({\varphi}(\tau)+{\varphi}_0)]-{{const{\times}{\beta}(\tau)}
\over{(1+{\gamma}_0^2{\beta}^2})}. \eqno(66)
$$

The value of $v_x$ given by (67) can be used to evaluate $v_y$ and $d$. We
can then compute $E(\tau)=(v_x^2+v_y^2+d^2)/2$, associated with the specific
spatial Fourier harmonic (SFH)
$$
E(\tau){\approx}{1 \over 2}{\left[C{\Omega}(\tau)+
{\left({const \over {\Omega}(\tau)}\right)^2}\right]}. \eqno(67)
$$

The energy has two main parts, the first one proportional to
$C{\Omega}(\tau)$ is due to the compressional perturbations, while the
second proportional to $const/{\Omega}(\tau)$ is of a purely vortical
nature. If the vortical part (mainly incompressible) is dominant, the
spectral energy of the SFH varies as
$$
E(\tau){\simeq}{1 \over {[1+{\gamma}_0^2({\beta}_0-R{\tau})^2]}}, \eqno(68)
$$
implying transient growth of SFH if ${\beta}_0=k_{y_1}/k_{x_1}>0$. The
energy has a maximum for ${\tau}={\tau}_*{\equiv}{\beta}_0/R$. This pattern
resembles the one obtained in the nonrelativistic case (Chagelishvili et al.
1994), but with the remarkable difference that the presence of ${\gamma}_0^2$
will lead to a much stronger growth for the superrelativistic case. In
particular, the ratio of the maximum value of the SFH spectral energy
(attained at the moment ${\tau}={\tau}_*$) to its initial value is:
$$
{{E_{max}}\over{E_{min}}}{\equiv}{{E({\tau}_*)}\over{E(0)}}=
1+{\gamma}_0^2{\beta}_0^2. \eqno(69)
$$
which, unlike its nonrelativistic counterpart, shows substantial increase
even
if ${\beta}_0{\le}1$. Thus relativity in conjunction with shear can induce
large transient growths for a wide range of $k_{y_1}$ and $k_{x_1}$.
Significant growth will be seen in regimes where there was virtually no
growth in the nonrelativistic case. And in the regimes of nonrelativistic
growth, the relativistic growth is ${\gamma}_0^2$ times larger.

For the compressional perturbations, ($C{\not=}0, const=0$) the energy change
takes the form
$$
E(\tau){\sim}[1+{\gamma}_0^2({\beta}_0-R{\tau})^2]^{1/2}. \eqno(70)
$$

If $k_{y_1}$ and $k_{x_1}$ are such that ${\beta}_0>0$, then for
$0<{\tau}<{\tau}_*$, the energy decreases and reaches its minimum at
${\tau}={\tau}_*$. For ${\tau}_*<{\tau}<{\infty}$, the SFH "emerges" into
the area of ${\bf k}$-space in which $k_y(\tau)k_{x_1}<0$, and begins to
grow. If initially ${\beta}_0<0$, the
perturbation energy monotonically increases. The perturbations grow up
effectively extracting the energy from the kinetic energy of the mean
(background) flow.

Like the case with vortical perturbations, the presence of
${\gamma}_0^2$ in (70) intensifies the process of the extraction
of the mean flow energy by the sound-type perturbations. The rate of
energy increase in the "growth area" (characterized by the derivative
${\partial}_{\tau}E$) is proportional to ${\gamma}_0^2$, and is quite fast
for large values of the Lorentz factor characterizing the mean flow.

\section{Conclusion}

In this paper we have investigated the novel features induced by relativity
in the acoustic phenomena by examining a
simple,  model parallel flow with  linear shear. The acoustic perturbations
are
excited in a background flow corresponding to an idealized, plane,
"shearing sheet" of a viscous fluid moving with relativistic speed. Principal
results of this study are:

In the shearless limit,
  when both the temperature of the medium  and the mean velocity of the
flow are strongly relativistic,  we find a noteworthy difference in
the frequencies of the sound waves propagating along   and
against the mean flow. Relativity also causes the violation
of the irrotationality of sound-type perturbations.

For the sheared system, we have first  derived the general system of equations
for the temporal evolution of the acoustic perturbations. We then analysed
in detail the particular case of the relativistic shear flow with a
nonrelativistic temperature. It was shown that the standard
shear-induced processes like the  nonexponential, algebraic temporal
evolution with transient growth of the vortical perturbations, and the ability
of the compressible (sound-type)
perturbations to extract   energy from the mean flow, persist in
relativistic dynamics. In fact, these processes become much more efficient;
the rate of a typical process is enhanced by
the square of the  Lorentz factor ${\gamma}_0^2$ which can be very large
for superrelativistic flows.

It is likely that the model flows discussed in this paper will have common
features with real  relativistic astrophysical flows. The possible
candidates are the flows in the relativistic regions
of accretion discs in quasars, jet-like outflows associated with Active
Galactic Nuclei (Granik \& Chapline 1996), pulsar winds, and  flows in
the  inner magnetospheres of pulsars (Lyubarsky 1995). In many
of these {\it real} astrophysical flows the
{\it special-relativistic} effects will be intricately
interlaced with the {\it general-relativistic}    effects evoked by strong
gravitational fields, and also with the  {\it electromagnetic}
effects (for plasma flows)
arising due to the existence of superstrong magnetic fields. We hope that the
methods used and the results obtained in the present study  will serve as
starting  points for some of the future
investigations of these interesting and complicated  relativistic  flows.

\section{Acknowledgements}

ADR's research was supported, in part, by International Science
Foundation (ISF) long-term research grant RVO 300. VIB's research was
supported, in part, by International Science Foundation (ISF) long-term
research grant KZ3200. ADR would like to thank George Chagelishvili,
George Machabeli and Otto Chkhetiani for valuable discussions.

\vskip 1.5cm
{\centerline{\Large \bf References}}
\vskip 0.5cm
Arons, J. \& Smith, D. F. 1979 Ap. J. {\bf 229}, 728.\\
Blandford, R. \& Rees, M.  Mon. Not. R. Astron. Soc. {\bf 169}, 395 (1974)\\
Chagelishvili, G. D., Rogava, A. D. \& Segal, I. N. 1994 Phys. Rev.~E
{\bf50}, 4283.\\
Chagelishvili, G. D., Rogava, A. D. \& Tsiklauri, D. G. 1996 Phys. Rev.~E
{\bf 53}, 6028.\\
Granik, A. \& Chapline, G. Phys. Fluids {\bf 8}, 269 (1996).\\
Lighthill, L. {\it Waves in Fluids} (Cambridge University Press,
Cambridge 1978).\\
Lyubarsky, Yu. E. 1995 Astrophys. Space Phys. {\bf 9}, 1.\\
Magnus, K. 1976, {\it Schwingungen} (B. G. Teubner, Stuttgart).\\
Rao, N. N., Shukla, P. K. \& Yu, M. Yu. 1990 Planet. Space Sci. {\bf 38},
543.\\
Rogava, A. D., Mahajan, S. M., \& Berezhiani, V. I. 1996
I.C.T.P. Internal Report (IC/IR/96/4) Phys. Plasmas (submitted).\\
Rogava, A. D., Chagelishvili, G. D. \& Berezhiani, V. I. 1996
I.C.T.P. Preprint (IC/96/37) (submitted to Phys. Rev. Lett).\\
Shukla, P. K. \& Silin, V. P. Phys. Scripta {\bf 45}, 508 (1992).\\
Sturrock, P.A. {\it Plasma Physics} (Cambridge, Cambridge University Press,
1994).
}

\end{document}